\documentclass[a4paper,11pt]{article}
\usepackage[utf8]{inputenc}

\usepackage{tabularx,booktabs,caption}
\usepackage{sidecap}
\usepackage{amsfonts}
\usepackage{amssymb}
\usepackage{graphicx}
\usepackage{amsmath}
\usepackage{enumerate}
\usepackage{color}
\usepackage{cite}
 \usepackage{multirow}
\usepackage{float}
\usepackage{caption}
\usepackage{cite}
\usepackage{epstopdf}
\RequirePackage{mathrsfs} \RequirePackage[sc]{mathpazo}
\RequirePackage{wasysym} \RequirePackage{setspace}\textheight=650pt
\title{ {\bf Phase Transition of Charged-AdS Black Holes and Quasinormal Modes : a Time Domain Analysis}}
\author{M. Chabab$^{1}$\footnote{mchabab@uca.ac.ma (Corresponding author)}, H. El Moumni$^{1,2}$\footnote{hasanelm@yahoo.fr}, S. Iraoui$^{1}$\footnote{s.iraoui@edu.uca.ma}, K. Masmar$^{1}$\footnote{karima.masmar@edu.uca.ac.ma}\\
	\\ 
	{\small $^{1}$ High Energy and Astrophysics Laboratory, Physics Department, FSSM, 
	}\\
         {\small Cadi Ayyad University, P.O.B. 2390 Marrakech, Morocco.
	}\\
	{\small $^{2}$ LMTI, Physics Departement, Faculty of Sciences,  Ibn Zohr University, Agadir, Morocco. }
}
\date{\today}

\begin{document}
 \maketitle

\begin{abstract}
 In this work we use the quasinormal mode  of a massless scalar perturbation to probe the phase transition of the charged-AdS black hole in time profile. The signature of the critical behavior of this black hole solution is detected in the isobaric process. This paper is a natural extension  of \cite{our7,Liu:art_basique} to the time domain analysis. More precisely, our study shows a clear signal  in term of the damping rate and the oscillation frequencies of the scalar field perturbation. We conclude that the quasinormal modes  can be an efficient tool to detect the signature of thermodynamic phase transition in the isobaric process far from the critical temperature, but fail to disclose this signature at the critical temperature.
 
 \noindent
 \\ \textbf{Keywords}:  Quasinormal modes,  $AdS$ black holes,  Phase transitions.

\end{abstract}

\section{Introduction}

The main motivation behind the study of black holes in the asymptotically AdS spacetime is provided by the gauge/gravity correspondence \cite{1h,2h}, which is a powerful tool applied to a broad variety of research areas like theoretical particle physics,  string theory, QCD, nuclear physics and condensed-matter physics \cite{3h,4h,5h,6h}. In this  context  \cite{1h,7h,8h,Mm}, the black hole is identified with an approximately thermal state in the field theory, and the decay of the test field corresponds to the decay of the perturbation of the state. In this way,  the understanding of asymptotically AdS black hole is a crucial  step  towards a more insight into the above conjecture.

Recently, a particular emphasis  has been dedicated   to the study of phases transitions of the black holes in AdS space \cite{KM, our} that consolidates the analogy between the critical Van der Waals gas behavior and the charged AdS black hole one \cite{our1,our2,our3,our4,our5,our6,our7}, using  several approaches based on mathematical methods. Especially  the quasinormal modes (QNM) \cite{our7, Liu:art_basique, indian} which has been proved efficient to disclose  the thermodynamic phase transition. More precisely, in \cite{our7, Liu:art_basique, indian} the authors foundd a drastic change in the slopes of the quasinormal frequencies in the small (SBH) and large (LBH) black holes near the critical point where the Van der Waals like thermodynamic phase transition occurs for different spacetime dimensions and gravity configuration using the frequencies domain analysis. Another important  motivation is related to new detection of the gravitational waves whose recent observation is an important landmark  in the gravitational waves astronomy \cite{mx1,mx28b}.

The aim of the work is to extend this technical method tothe time-domain analysis and show a correspondence between the thermodynamical phase transition and the time evolution of the wave function. 

The layout of the letter is as follows: in section 2 we briefly review some aspects of the phase transition in the extended phase space in the temperature-horizon radius plan. In section 3 we present the master equation of the evolution of a massless scalar field in the four dimensional RN-AdS background as well as the numerical procedures used to solve Section $4$ is devoted to establish a link between the damping rate, the oscillation number of the scalar field and the small/large black hole phase transition and to show that the behavior discussed in our previous work \cite{our7} is well reproduced. In the last section, we will draw our conclusions.

%%%%%%%%%%%%%%%%%%%%%%%%%%%%%%%%%%%%%%%% 
\section{Critical behavior of $\text{RN-AdS}_4$ black holes}
Let us recall some aspects of the system to be studied. Reissner Nordstrom AdS black hole is solution of Einstein-Maxwell theory with negative cosmological constant in four dimensions which described by the following action
\begin{equation}\label{actem}
	I_{EM}=-\frac{1}{16\pi G_4}\int_M d^4x\sqrt{-g}\Bigl(\mathcal{R}-F^2+\frac{6}{R^2}\Bigr)\,.
\end{equation} 
where $\mathcal{R}$ is the Ricci scalar, $F_{\mu\nu}$ is the strength of electromagnetic field and $R$ is the curvature radius of AdS space. We have assumed the universal gravitational constant $G_4=~1$. Solving Einstein equations yields to the following the static spherical metric  
\begin{equation}\label{HDRN}
ds^2 = -f dt^2 + \frac{dr^2}{f} + r^{2} d\omega_{2}^2,
\end{equation}
with, 
\begin{equation}\label{metfunction}
f = 1 - \frac{2M}{r} + \frac{Q^2}{r^{2}} + \frac{r^2}{R^2}\,,
\end{equation}
and $d\omega_2^2$ stands for the metric of a four dimensional unit sphere. The two parameters $M$ and $Q$ are the  mass   and the charge of the black hole respectively. The Hawking temperature  reads as
\begin{equation}\label{T4}
T = \left.\frac{f'(r)}{4 \pi}\right|_{r=r_H}
= \frac{1}{4 \pi r_{H}} \Bigl(1 - \frac{Q^2}{r_H^{2}} +\frac{3 r_H^2}{R^2} \Bigr)\,,
\end{equation}
where  the position of the black hole event horizon $r_H$ is  determined by solving the equation $f(r)|_{r=r_H}=0$ and choosing the largest real positive root. The electric potential $\Phi$ measured at infinity with respect to the horizon while the black hole entropy $S$ are given by the following form
\begin{eqnarray}\label{S4} 
\Phi&=&\frac{Q}{r_H}\,,\\
S &=& \pi r_H^2\,.
\end{eqnarray}

Now, we reconsider  the extended phase space by defining a thermodynamical pressure proportional to cosmological constant and its corresponding conjugate quantity as the volume, using the following equation
\begin{equation}
P = - \frac{\Lambda}{8 \pi} =\frac{1}{8 \pi R^2}\,, 
\end{equation}

% All these quantities satisfy the Smarr formula \cite{KM}
% \begin{equation}
% M=TS+\Phi Q-2 VP\,,
% \end{equation}  
In this context the mass is identified with the enthalpy \cite{a1,KM}, the first law of black hole thermodynamics becomes, 
\begin{equation}
dM=TdS+\Phi dQ+VdP\,,
\end{equation}
where the thermodynamic volume can be defined as
\begin{equation}
 V=\left.\frac{\partial M}{\partial P}\right|_{S,Q}.
\end{equation}

To the Gibbs free energy $G=M-TS$,  for fixed charge, it reads as \cite{4}
\begin{equation}
\label{GibbsQ}
G=G(P,T)
=\! \frac{1 }{4} \biggl(\!r_H - \frac{8\pi Pr_H^{3}}{3}
+ \frac{3Q^2}{r_H}\biggl).\ \ 
\end{equation}

\begin{figure}[!ht]
\begin{center}
	\includegraphics[height=5.cm]{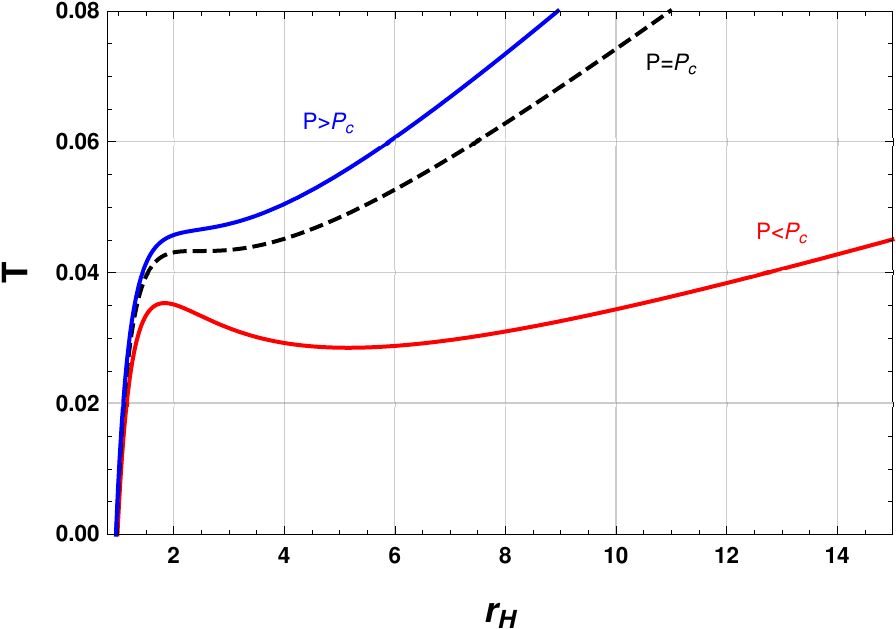}\hspace*{4em}\includegraphics[height=5.cm]{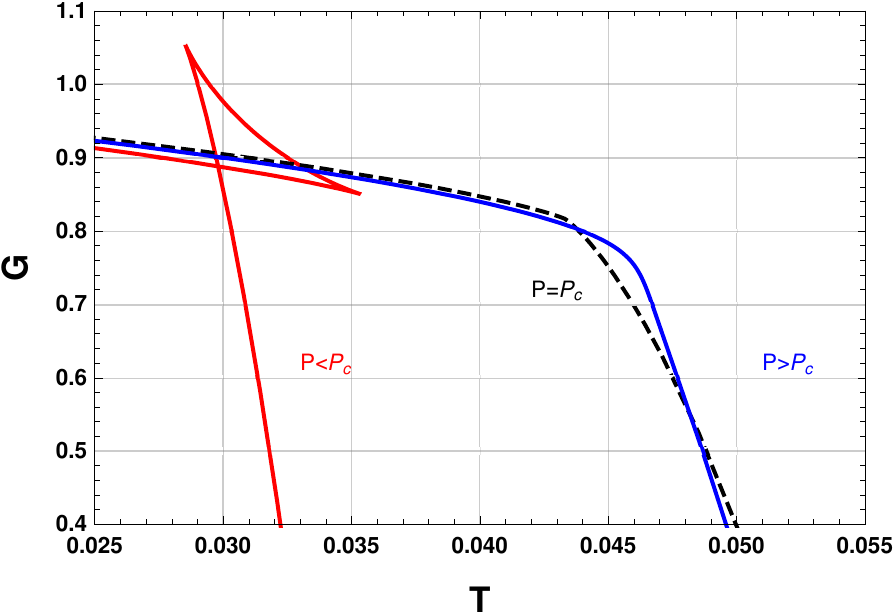} 
\vspace*{-.2cm} \caption{\footnotesize{\bf Left:} Isobaric curves of  RN-AdS black hole with different pressure. {\bf Right:} The Gibbs free energy function of the temperature. For each panel from bottom to top the corresponding pressure are $0.4 P_c$ (red), $P_c$ (Dashed Black), $1.2 P_c$ (Blue)\label{fig1}
}
\end{center}
\end{figure}

Now, having calculated the relevant thermodynamic quantities, we turn now to the analysis of the corresponding phase transition. For this, we plot in figure \ref{fig1}  the variation of the Hawking temperature and the Gibbs free energy as a function of the horizon radius given by Eq. \eqref{T4} and Eq. \eqref{GibbsQ}. We set   the charge $Q$ to $1$ for all the rest of this letter.

As discussed in \cite{KM,our}, there exists some critical behavior. In particular the first order phase transition in the $P-V$ diagram signaled by the “swallow tail” shown in the right panel of figure \ref{fig1}. This behavior is similar  to the Van der Waals one. We perform a similar calculation on the $T-r_H$ plane by fixing the pressure. The coordinates of the critical point can be obtained by solving the following system of equations
\begin{equation}
\frac{\partial^2 T}{\partial r_H^2}= \frac{\partial T}{\partial r_H}=0,
\end{equation}
which give rise to the following critical pressure, critical radius  and the critical temperature 
\begin{equation}
 P_c=\frac{1}{96 \pi},\quad r_{H_{c}}=\sqrt{6}, \quad \text{and }\; T_c= \frac{1}{3\sqrt{6}\pi}.
\end{equation}

% 
% 
% Various values of
% $P$ is employed to plot the behavior of the temperature and the Gibss free energy. The
% % Blue curve corresponds to $P>P_c$ which a monotonic function corresponds to the Schwarzschild-AdS black
% % hole. There, the temperature has a minimum (similar to the
% % case described
% % above).
% When $P$
% is increased, the temperature lost her monotony and present an inflection point 
% 
% 
% 
% has two turning points (minima and a
% maxima). Further increasing the temperature brings these two tu
% rning points to one
% point leading to an inflection point. After that the curve does not ha
% ve any turning
% points. The critical charge
% $Q_c$ is the value of $Q$
% at which the 
% $T\;vs\;r_h$ curve has an inflection point. At this point,

After having briefly introduced the main thermodynamical quantities and  related  phase transition, we will study in the next section the late–time decay of a scalar perturbation around a charged AdS black holes in four dimensional spacetime.

%%%%%%%%%%%%%%%%%%%%%%%%%%%%%%%%%%%%%%%%%%%%%%%%%
\section{Master equation and numerical method}
We start by considering  the evolution of a massless scalar field in charged AdS black hole  background, satisfying  the Klein-Gordon differential equation
\begin{equation}
	\frac{1}{\sqrt{-g}}\partial_{\mu}\left(\sqrt{-g} g^{\mu\nu}\partial_{\nu}\Phi\right)=0,
\end{equation}
then,we decompose  the scalar  field as
\begin{equation}
	\Phi=\sum_{\ell ,m} r^{-1}\psi(t,r)Y_{\ell ,m}(\theta , \phi),
\end{equation}
to separate the radial and  angular variables. The radial equation is giving by  \cite{Zhu:2001vi,Li:2016kws}
\begin{equation}\label{kg1}
	-\frac{\partial^{2}\psi}{\partial^{2}t}+f\frac{\partial}{\partial r}\left(f\frac{\partial \psi}{\partial r}\right)=V_{\ell} \psi,
\end{equation}
where the effective potential $V_{\ell}$ has the form \cite{Toshmatov:2016bsb,Morgan:2009vg} 
%\begin{equation}
%	V_{\ell}=\left[\frac{\ell(\ell+d-3)}{r^{2}}-\frac{(2-d)(4-d)}{4r^{2}}f-\frac{(2-d)}{2r}\frac{\partial f}{\partial r}\right]f,
%\end{equation}
\begin{equation}
V_{\ell}=\left[\frac{\ell(\ell+1)}{r^{2}}+\frac{1}{r}\frac{\partial f}{\partial r}\right]f,
\end{equation}
where $\ell$ is the angular quantum number. In this letter we will focus only  on the case $\ell =0$ and  we     tortoise coordinate  defined by $r^{*}=\int \frac{dr}{f}$ (up to an arbitrary constant). If we take  $r^{*}\left(r\rightarrow\infty\right)~=~0$, the analytic form of $r^*$ in function of $r$ is giving by \cite{tortoise,Zhu:2014sya,Wang:2000dt}
\begin{multline}\label{tort} 
	r^{*}=\frac{\log
		\left(r-r_H\right)}{2 \kappa _H}-\frac{\log \left(r-r_C\right)}{2 \kappa _C}-A \log \left(T_{2} r+T_{1}+r^2\right)\\ +\frac{2 (A T_{2}+B)}{\sqrt{4 T_{1}-T_{2}^2}}\left[\tan ^{-1}\left(\frac{T_{2}+2 r}{\sqrt{4 T_{1}-T_{2}^2}}\right)-\frac{\pi }{2}\right], 
\end{multline}
with\\
$T_{1}=r_H^2+r_C^2+r_H r_C+R^2$,\ \ $\kappa _H=\frac{\left(r_H-r_C\right) \left(3 r_H^2+r_C^2+2 r_H r_C+R^2\right)}{2 r_H^2 R^2}$,\ \  $A=\frac{ R^2 \left(r_H+r_C\right) \left(r_H^2+r_C^2+2 r_H r_C+R^2\right)}{2 \left(3 r_H^2+r_C^2+2 r_H r_C+R^2\right) \left(r_H^2+3 r_C^2+2 r_H r_C+R^2\right)}$,\\ \\
$T_{2}=r_{H}+r_{C}$,\hspace*{5.5em} $\kappa _C=\frac{\left(r_H-r_C\right) \left(r_H^2+3 r_C^2+2 r_H r_C+R^2\right)}{2 r_C^2 R^2}$\ \ \; and\;
$B=\frac{R^2 \left(r_H^2+r_C^2+R^2\right) \left(r_H^2+r_C^2+r_H r_C+R^2\right)}{\left(3 r_H^2+r_C^2+2 r_H r_C+R^2\right) \left(r_H^2+3 r_C^2+2 r_H r_C+R^2\right)}$.\\

$r_C$ and $r_H$ are Cauchy and event horizons respectively.

Therfore by using the tortoise coordinate, the Eq. \eqref{kg1} reduced to \cite{Konoplya:2011qq,Santos:Shwaezshild,Zhu:2001vi}
\begin{equation}\label{kg2}
	-\frac{\partial^{2}\psi}{\partial^{2}t}+\frac{\partial^{2} \psi}{\partial^{2} r^{*}}=V_{\ell} \psi.
\end{equation}
In order to  integrate the Eq. \eqref{kg1} we use the numerical method developed  by Gundlach, Price and Pullin \cite{Gundlach:1993tp}. We start by  introducing the null coordinates $u=t-r^{*}$ and $v=t+r^{*}$ to simplify  the Eq. \eqref{kg2} which can be rewritten as two dimensional wave equation
\begin{equation}\label{kg3}
	-4 \frac{\partial^{2}\psi\left(u,v\right)}{\partial u\partial v}=V_{\ell}\left(r\right) \psi\left(u,v\right).
\end{equation}
Now the problem is reduced  to the numerical integration of the Eq. \eqref{kg3}, the values of the wave function $\psi$ can be found via the discretization method \cite{Gundlach:1993tp,Zhu:2001vi,Santos:Shwaezshild,tortoise},  which we  upon call Taylor’s theorem
\begin{equation}\label{kgnum}
\psi(N)=\psi(W)+\psi(E)-\frac{1}{8}\delta v\delta u V_{\ell}(\frac{v-u}{2}) \left(\psi(W)+\psi(E)\right)\\ -\psi(S)+\mathcal{O}(\delta v\delta u),
\end{equation}	
where $S=(u,v)$, $W=(u+\delta u,v)$, $E=(u,v+\delta v)$ and $N=(u+\delta u,v+\delta v)$ form a null grid in the $v-u$ plane with horizontal step $\delta v$ and vertical step $\delta u$. In order to calculate the effective potential we need to invert numerically $r^{*}\left(r\right)=\frac{v-u}{2}$ to $r$ by using  Eq. \eqref{tort}.

\begin{SCfigure}[][!ht]
			\includegraphics[width=9cm,height=6.cm]{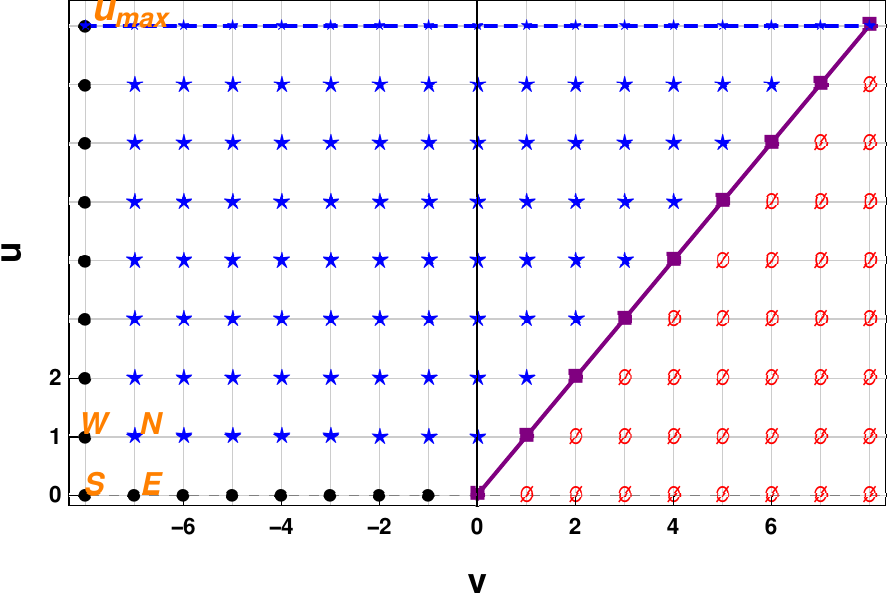}
		\vspace*{-.4cm}\caption{\footnotesize Diagram of the numerical grid\label{figgrille}}
\end{SCfigure}

From Eq. \eqref{tort} we see that $r^*$ is restricted within the range $-\infty<r^{*}<0$,  which  means that the only physical region in our problem is given by negative value of $r^{*}$ and  only the case $v\leq u$ in our calculation has to be considered. In the  purple line of figure \ref{figgrille}, where $v=u$ we set $\psi=0$ since  $r=\infty$ and $V_{\ell}$ diverge along this line  \cite{Zhu:2001vi}. Our objectif is  evaluate the time evolution of the wave near the horizon, for this, we  choose a sufficiently large $u_{max}$ by keeping $u_{max}>>v$, where $u_{max}$ is the maximum value of $u$ on the numerical grid, i.e when $u_{max}\rightarrow \infty$ the value of $r^{*} \rightarrow -\infty$, so $r$ tends to the horizon radius.

In  figure \ref{figgrille}, the black   spots represent the initial grid points, the blue  stars represent the grid points to be calculated, the red empty sets represent the forbidden region. The grid shows the iterative calculation  up $u=u_{max}$  corresponding to the dashed blue line.

As  in \cite{Santos:Shwaezshild,Gundlach:1993tp}, to solve the equation of perturbation   under the assumption that the wave function is insensitive to the choice of the initial data, we impose the following conditions:
\begin{itemize}
	\item Constant data on $v=v_0$, $\psi(u,v_{0})=\psi_{0} $.
	\item Gaussian distribution $\psi(u=u_{0},v)=A \exp\left(-\frac{\left(v-v_{c}\right)}{2 \sigma^{2}}\right)$  on $u=u_0$.\end{itemize}
	
	In addition, the following parameters are fixed, $\psi_{0}=0$, $\sigma=3$, $A=1$, $v_{c}=0$, $u_{0}=0$, $v_{0}=-10$ and  $\delta u=\delta v$.

After presenting the essential of the numerical method we turn our attention in the next section to the numerical  simulation results, and the signature of the thermodynamical phase  transitions in these results. 

%%%%%%%%%%%%%%%%%%%%%%%%%%%%%%%%%%%%%%%%%%%%%%
\section{Time domain profile and isobaric phase transition} 
In our previous work \cite{our7} \footnote{See also \cite{Liu:art_basique}}   we found that the quasinormal mode frequencies  are able to probe the small/large black hole phase transition  in the isobaric process. In this section we will check this result by considering a time domain analysis.
%We fix the pressure at  $P=0.00132$ ($P=0.4P_c$) which corresponds to the red line in figure \ref{fig1}.

\begin{figure}[!ht]
\begin{center}
	\includegraphics[width=7.8cm,height=5.5cm]{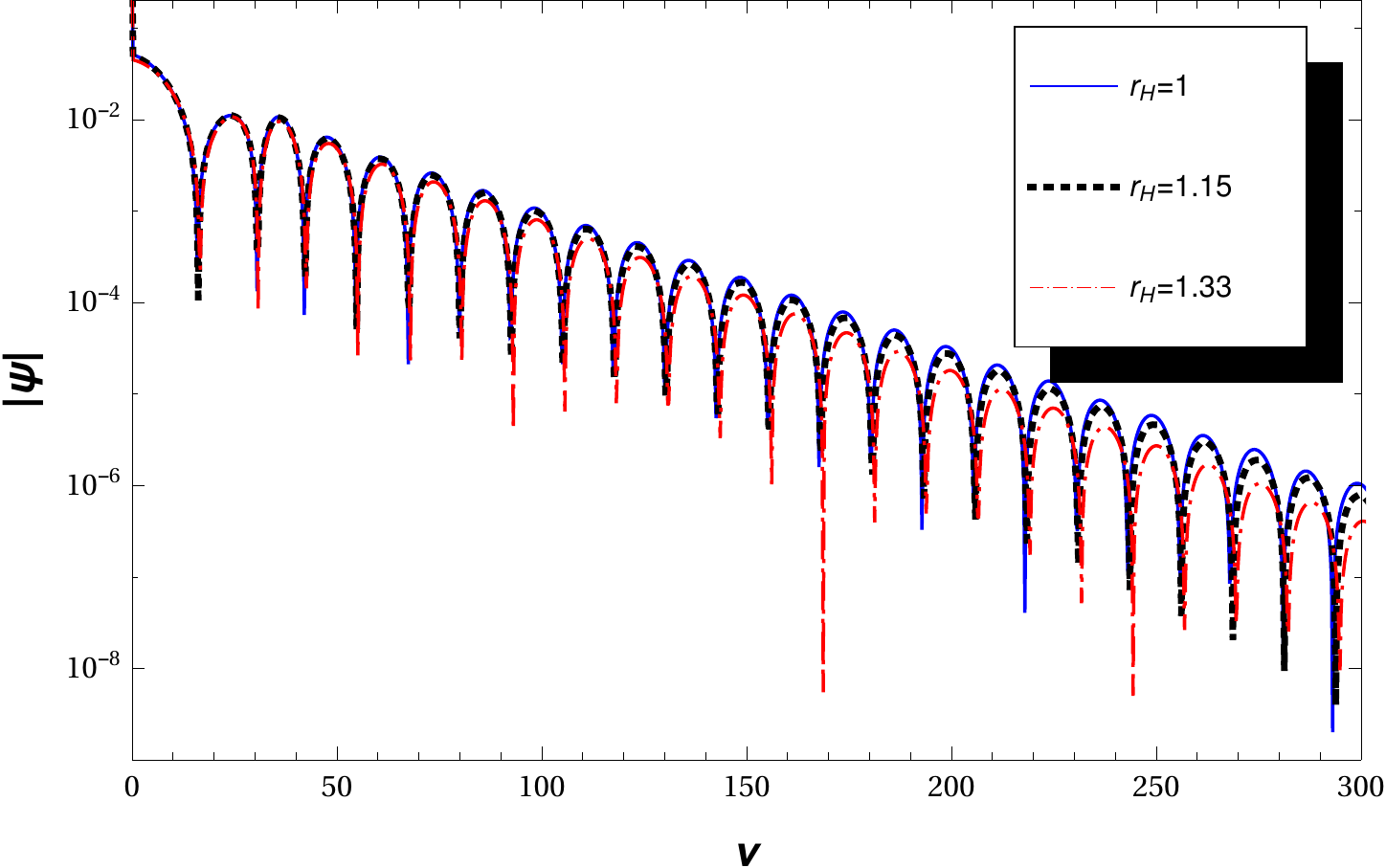}\hspace*{1em}\includegraphics[width=7.8cm,height=5.5cm]{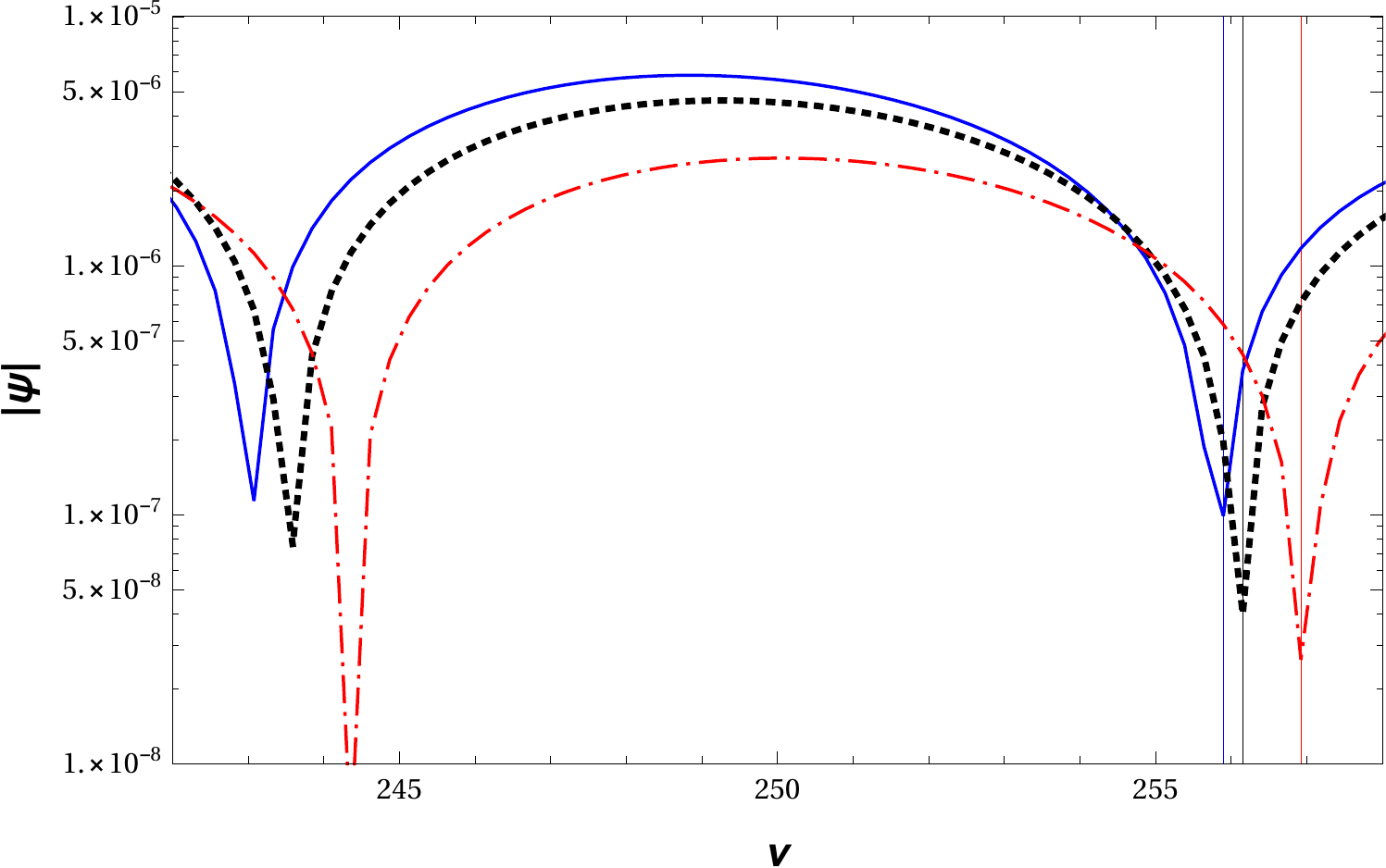} 
\vspace*{-.3cm} \caption{\footnotesize Left panel represent the time evolution of a scalar perturbation at the small $\text{AdS}_4$ black hole horizons (semi-log graph of $|\psi|$) for $r_H=1,1.15\ \text{and}\ 1.33 $. Right panel is a zoom of the left one. The pressure is fixed at $P=0.4P_c$\label{figtransitionsmall}
}
\end{center}
\end{figure}

\begin{SCfigure}[][!ht]
\includegraphics[width=9cm,height=6cm]{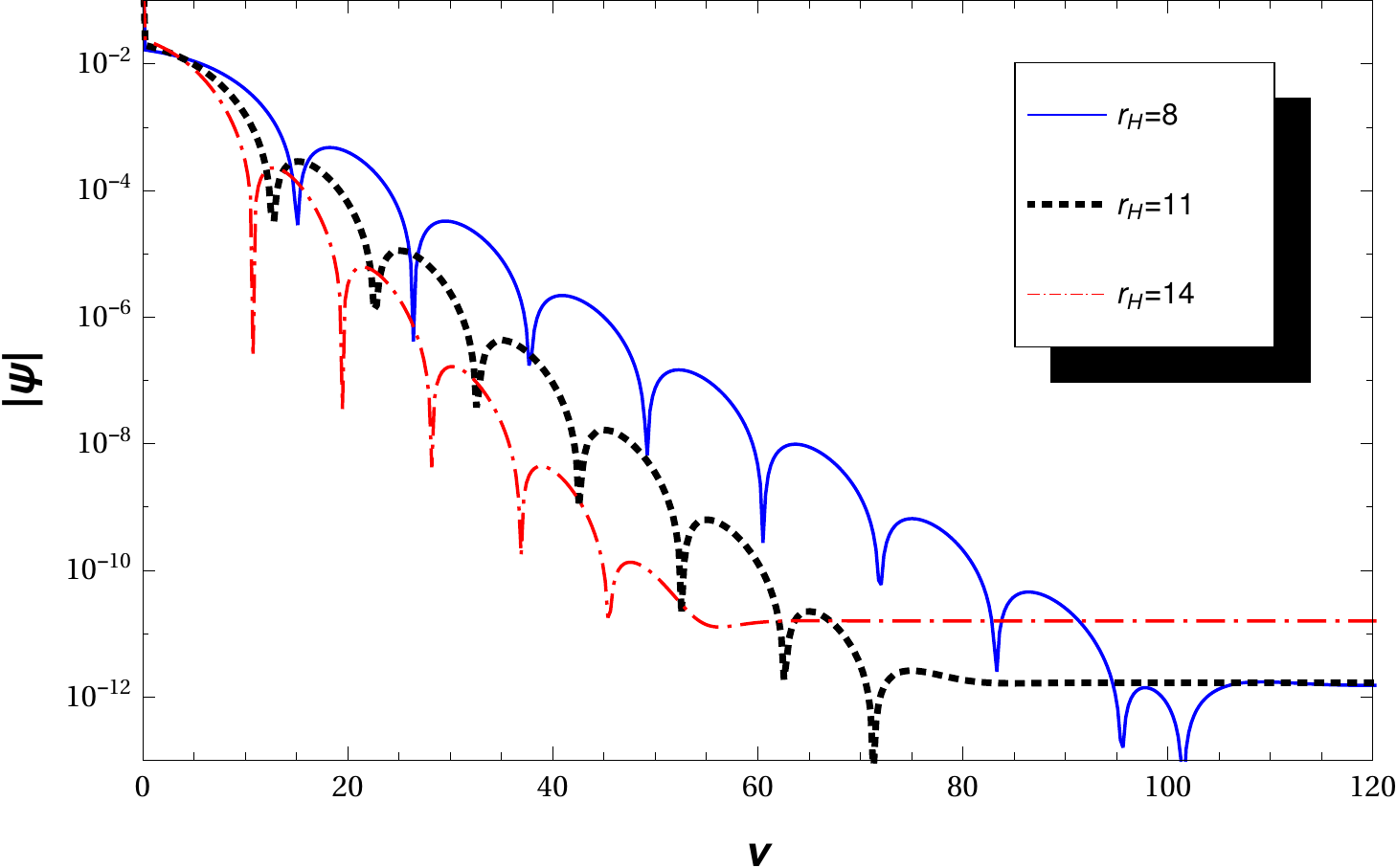}
\vspace*{-.3cm} \caption{\footnotesize Time evolution of a scalar perturbation at the large $\text{AdS}_4$ black hole horizons (semi-log graph of $|\psi|$) for $r_H=8,11\ \text{and}\ 14 $. The pressure is fixed at $P=0.4P_c$\label{figtransitionlarge}}
\end{SCfigure}

To this end we computed the time evolution of the perturbation for 3 values of $r_H$ at the BH horizon by fixing the pressure at $P=0.00132$ ($P=0.4P_c$). The figure \ref{figtransitionsmall} and \ref{figtransitionlarge} correspond to the small and large black hole respectively. We found that for both SBH and LBH, the quasinormal ringing is quickly fading out with the increasing size of the BH. 

\begin{SCfigure}[][!ht]
	\includegraphics[width=9cm,height=6cm]{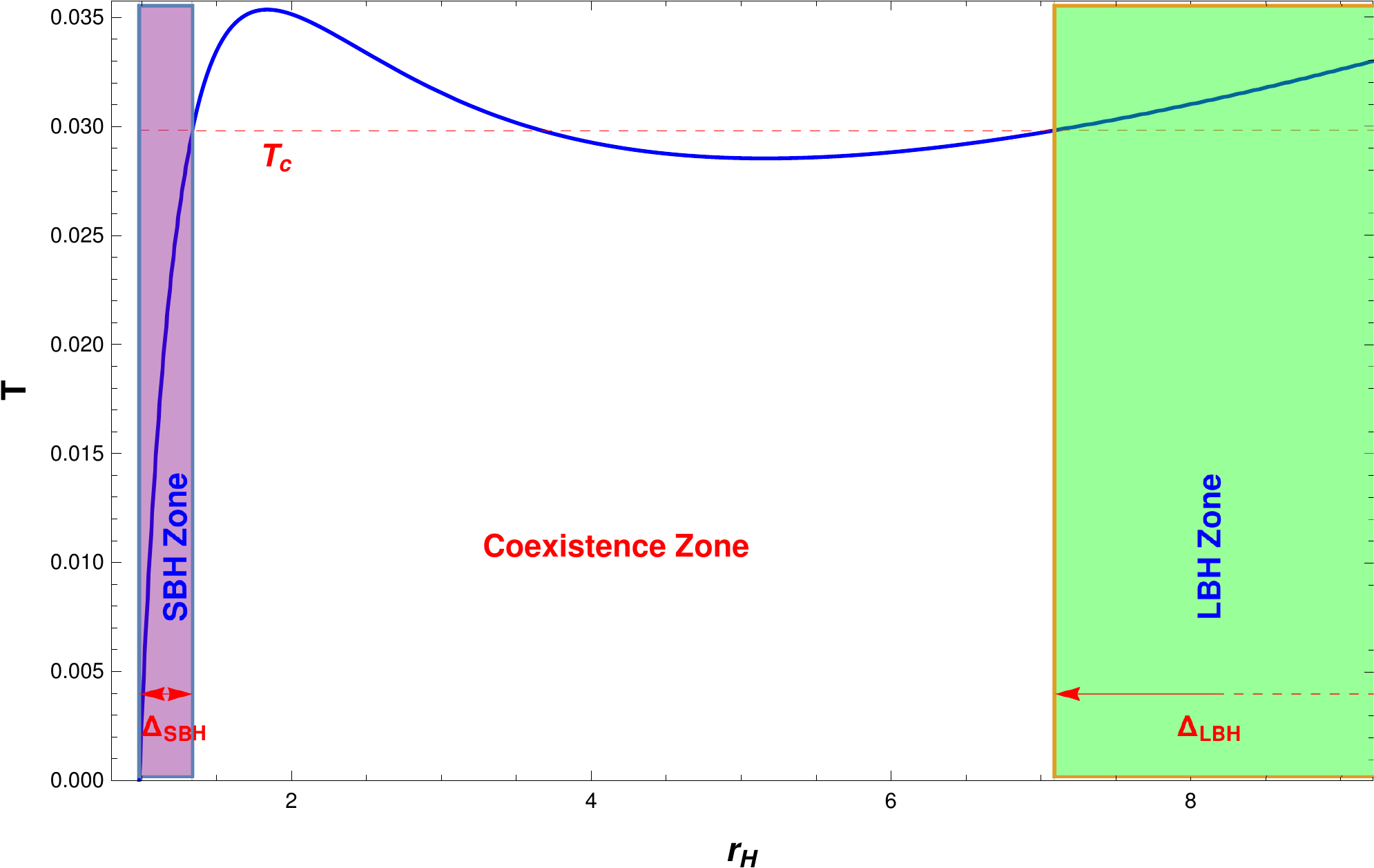}
	\vspace*{-.3cm} \caption{\footnotesize Isobaric $T-r_H$ diagram of charged $\text{AdS}_4$ black hole with $P=0.4P_c$. In general the range of $r_H$ in small black hole zone is very reduced $\Delta_{SBH}=0.353903$, contrariwise it is infinite in the large BH zone $\Delta_{LBH}=\infty$ \label{rangesmall}}
\end{SCfigure}

Form figure \ref{figtransitionsmall} we can see  that for the small black hole, when the horizon radius grows, the perturbation decays faster and the oscillation period increases. It is interesting to note that the increasing of oscillation frequencies  is very slow, hence we cannot clearly distinguish the ringing of time evolution for each $r_H$. Indeed if we choose a pressure value far from the critical one, we will get very reduced range of the possible values of $r_H$ restricted by the extremal radius. One corresponds to  $T=0$, while the other represents the maximum radius of SBH  where $T=T_c$. 
As illustrative example with
$P=0.4P_c$ one find $r_H(T=0)=0.984235$ and $r_{H_{SBH}}(T_c)=1.33814$. On the other hand whatever the pressure  of large black holes there is an infinity of possible values of $r_H$, as shown in figure \ref{rangesmall}.

From figure \ref{figtransitionlarge}, for $\text{AdS}_4$ large black hole phase, the damping rate is monotonically  increasing with the horizon radius $r_H$, but the oscillation period decrease with the increasing of $r_H$. The decay of the perturbation becomes faster for both small and large black holes with the growing of $r_H$, since the absorption ability of the black hole is enforced  \cite{Liu:art_basique}.

To make connexion with our previous work \cite{our7}, we briefly present our results  from frequency domain analysis.

\begin{SCfigure}[][!ht]
\includegraphics[width=5cm,height=3cm]{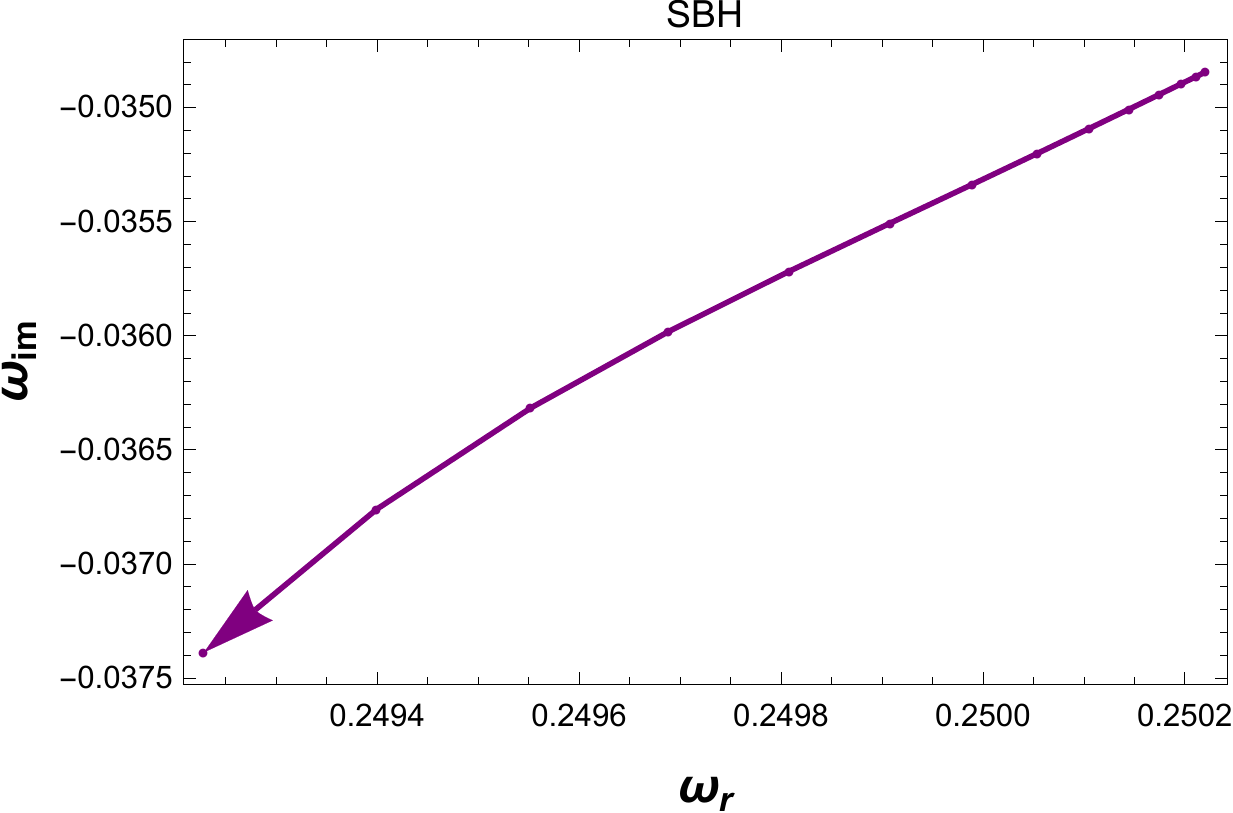}\includegraphics[width=5cm,height=3cm]{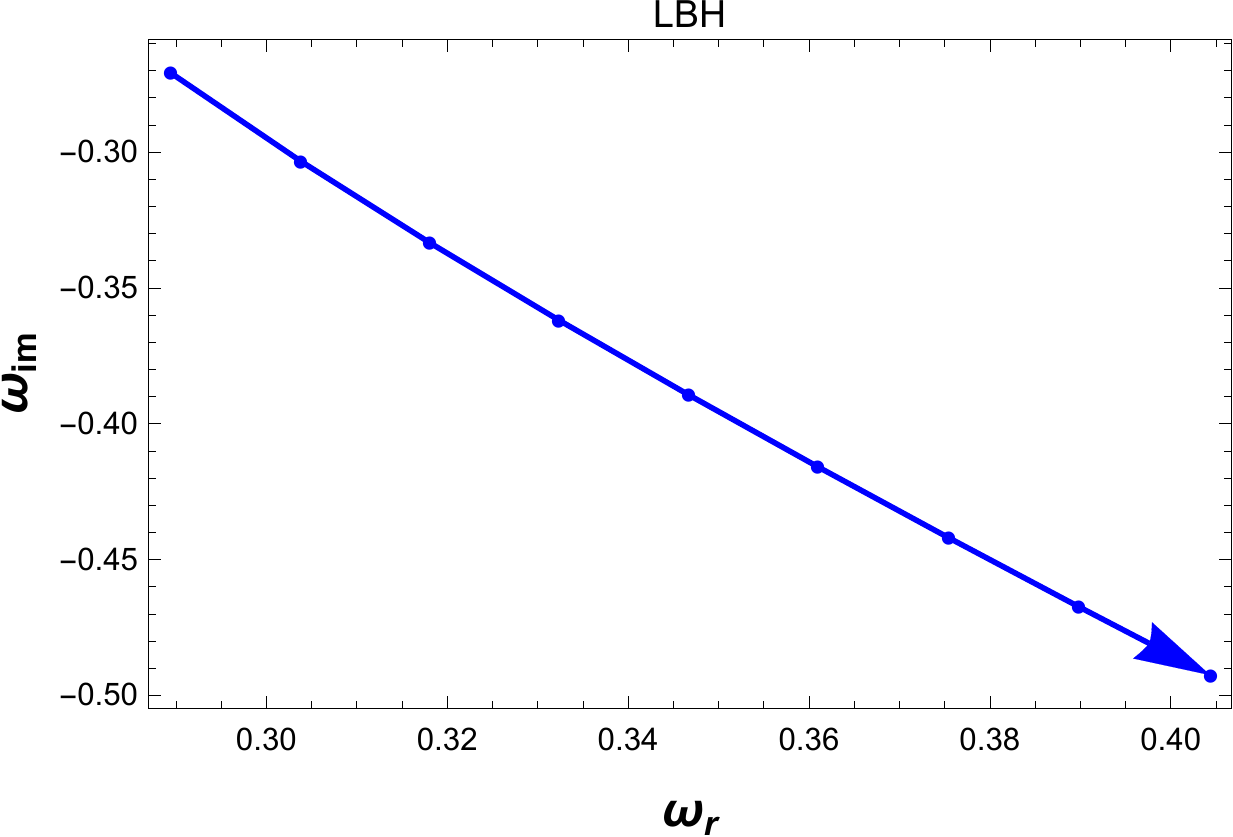}
\vspace*{-.3cm}\captionof{figure}{\footnotesize The  behavior of the quasinormal modes for small and large black holes in the complex-$\omega$ plane. 
Increase of the black hole size is shown by the arrows. The pressure is fixed at $P=0.4P_c$\label{figfrequencies}}
\end{SCfigure}

Table \ref{tab1} lists the frequencies of the quasinormal modes  of the massless scalar perturbation around small and large black holes for the first order phase transition where the pressure is fixed at the value $P=0.4P_c$. In figure \ref{figfrequencies} we illustrate the  quasinormal frequencies for SBH and LBH phases in $\omega_{r}-\omega_{im}$ plane, where the purple (blue) dots correspond to the results above (below) the horizontal line in table \ref{tab1} respectively.

\begin{table}[!htp]
\begin{tabular*}{\linewidth}{p{0.8cm}@{}l@{\extracolsep{\fill}}lll@{}} 
			\cline{2-5}
			 & $r_{H}$ & $T$ &$\omega_{r}$  & $\omega_{im}$ \\%& 	\centering{$T$} & \centering{$r_{H}$} & \centering{$\omega_{r}$}  & \hspace{9pt}$\omega_{im}$\\ 
			\cline{2-5}
			%\multicolumn{4}{|c}{$d=4$; $P=\frac{1}{96 \pi}$ ; $T_{c}=\frac{1}{3\sqrt{6}\pi}$}}&  	\hline				 
				    	 	 \multirow{3}*{\rotatebox{90}{\textbf{SBH}}}
				    	 	& 1. & 0.00265258 & 0.250222 & -0.0348405 \\
				    	 	& 1.15 & 0.0199248 & 0.249863 & -0.0356006 \\
				    	 	& 1.33 & 0.0295358 & 0.249177 & -0.03761 \\
				    	 	\cline{2-5} \multirow{3}*{\rotatebox{90}{\textbf{LBH}}}& 8. & 0.0310124 & 0.276141 & -0.237695 \\
				    	 	& 11. & 0.0363529 & 0.31498 & -0.327022 \\
				    	 	& 14. & 0.0427913 & 0.360997 & -0.415881 \\
				\cline{2-5} 
\end{tabular*}
\captionof{table}{\footnotesize  The quasinormal frequencies of the massless scalar perturbation as a function of  the black holes temperature. The upper part, above the horizontal lines is for the small black hole phase while the lower part is for the large one. The pressure is fixed at $P=0.4P_c$\label{tab1}}
\end{table}

From table \ref{tab1}, we see that for small black hole process, the real part of the QNMs frequencies varies slightly, while the absolute value of the imaginary part decreases while in the large black hole phase the real part as well as the absolute value of the imaginary part of quasinormal frequencies increase. These observation can be driven from figure \ref{figfrequencies} where we see different slopes of the quasinormal frequencies in the massless scalar perturbations with different phases of the small and large black holes. 

According to the discussion above, one can see that our results in the present work are in good agreement with \cite{our7,Liu:art_basique}.As a summary: The damping rates for both SBH and LBH have the same behavior, the oscillation period  decreases for the SBH while it increases for LBH. This result presents a consistent picture with those given by the quasi normal frequencies calculated in \cite{our7,Liu:art_basique,indian} which testify that the time domain analysis can be a useful tool to uncover a signature of  the first order phase transition between SBH and LBH. 

\begin{SCfigure}[][!ht]
	\includegraphics[width=5cm,height=3cm]{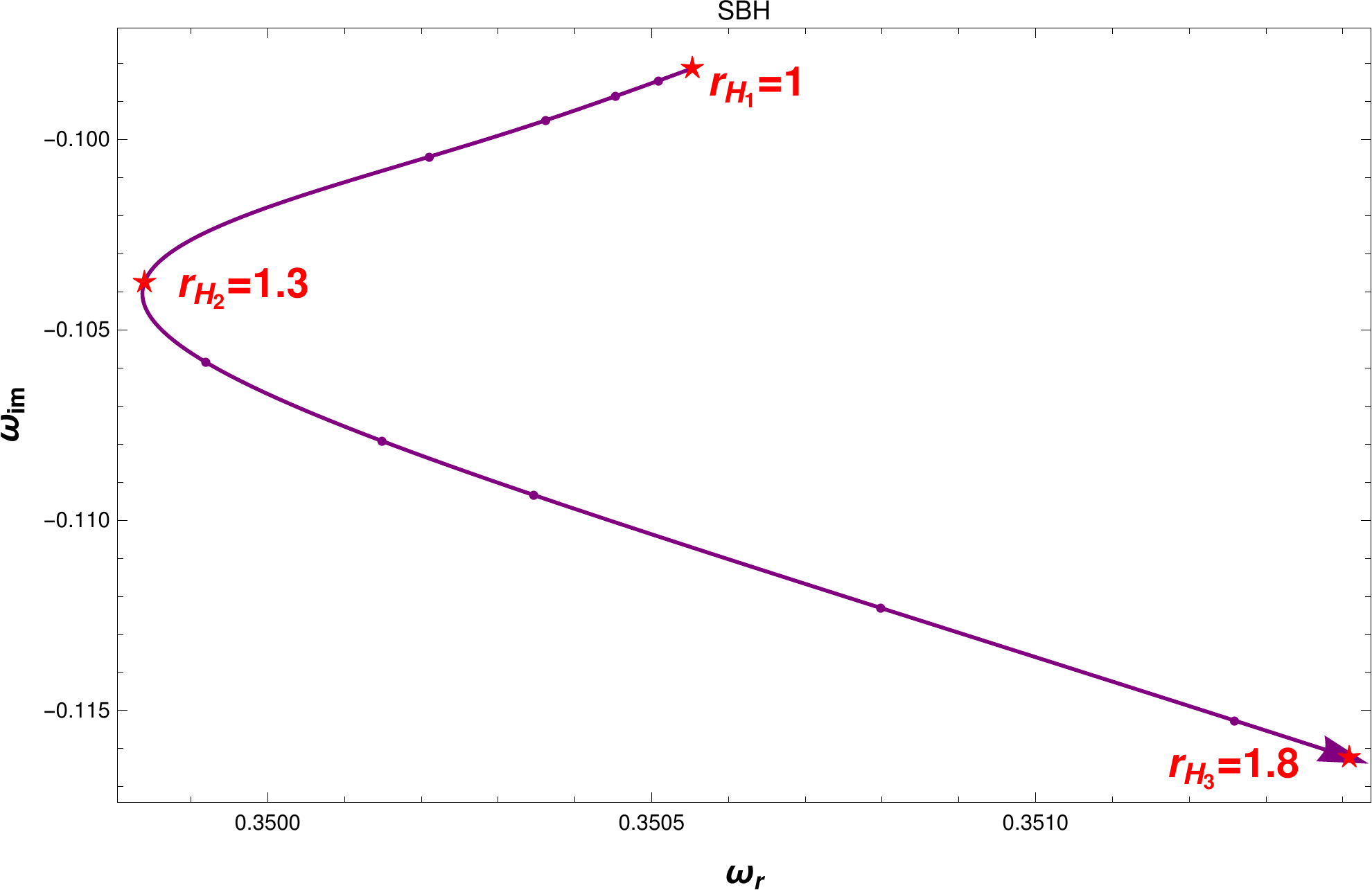}\hspace*{1em}\includegraphics[width=5cm,height=3cm]{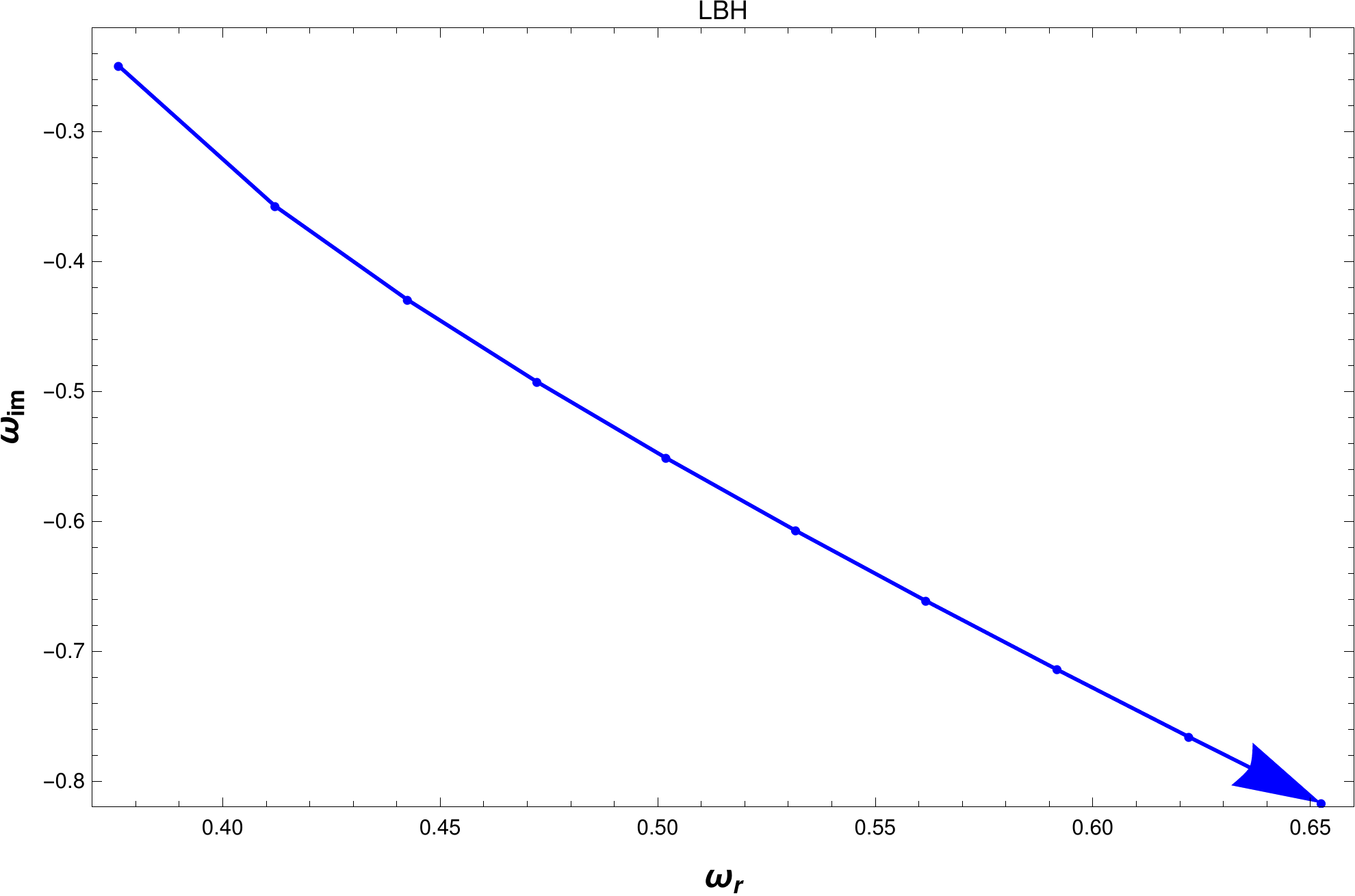}
	\vspace*{-.3cm}\captionof{figure}{\footnotesize The  behavior of the quasinormal modes for small and large black holes in the complex-$\omega$ plane. Increase of the black hole size is shown by the arrows. The pressure is fixed at $P=0.9P_c$\label{freqrappcriti}}
\end{SCfigure}

Besides, in our previous work \cite{our7} we also found that the critical ratio defined by $\chi=\frac{P}{P_c}$ affect the behavior of QNMs in the complex-$\omega$ plan, especially for small black hole. Probing the phase transition by the quasinormal modes in the frequency domain depends on the value of $\chi$. From figure \ref{freqrappcriti}, we see that for $\chi=0.9$ the $\omega$-complex plane does not show different slope in SBH and LBH phases. The  change of slope (from the positive to negative slope)  appears in the small black holes below the coexistence point. By approaching more and more  the critical pressure, the QNMs become less effective to probe the phase transition and the  slope change disappears at $\chi=1$. 

Apart a preliminary numerical analysis proposed by Subhash Mahapatra  in \cite{indian}, we do not have yet an analytic formula or a powerful condition governing the slope in $\omega_{r}-\omega_{im}$ plane that allows QNMs to effectively probe the phase transition.  The difficulty comes from from the master equation \eqref{kg1} which only depends  on  the parameters $P$ and $r_H$, regardless whether the black hole is large or small .

\begin{figure}[!ht]
\begin{center}
		\includegraphics[width=7.8cm,height=5.5cm]{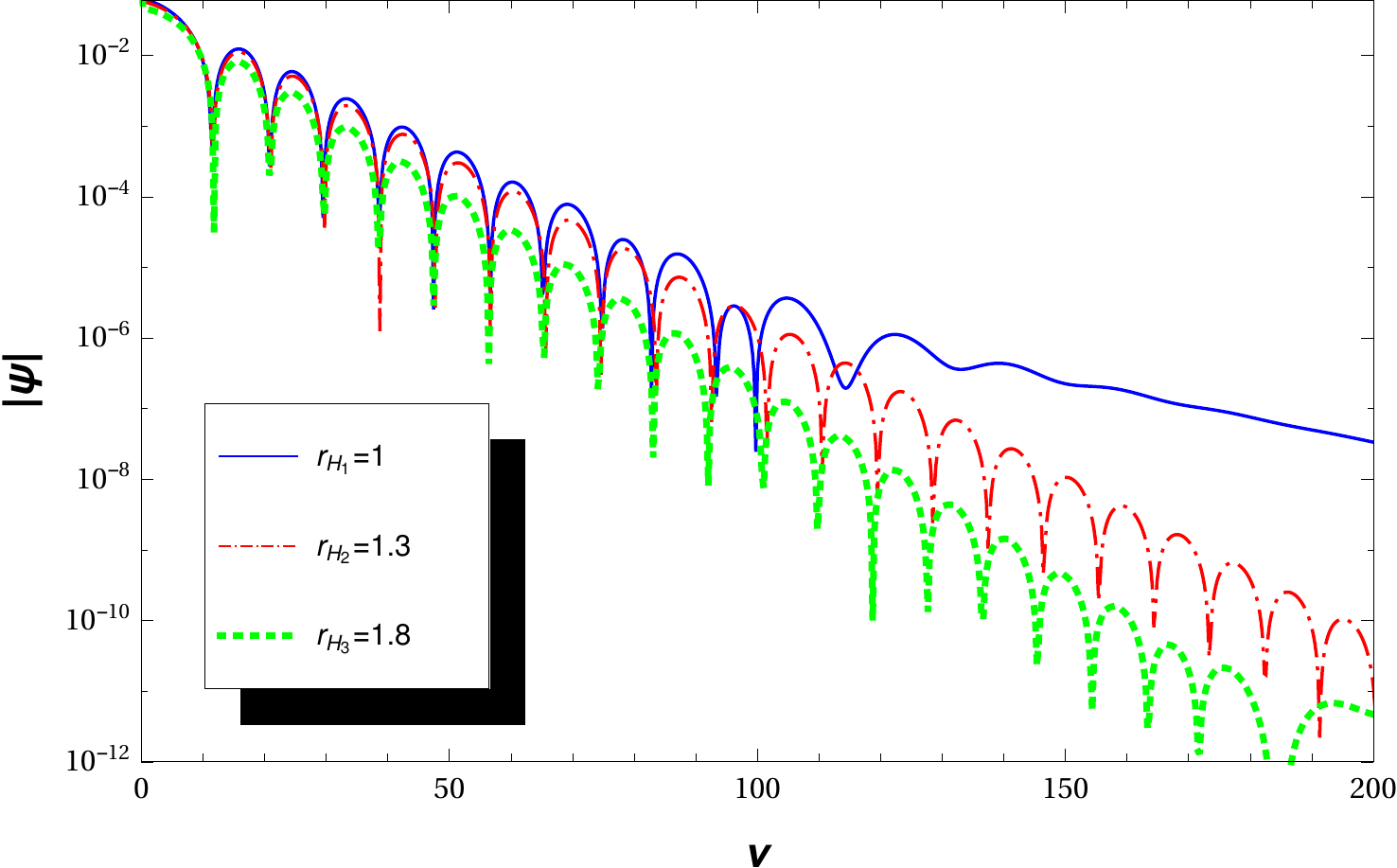}\hspace*{1em}\includegraphics[width=7.8cm,height=5.5cm]{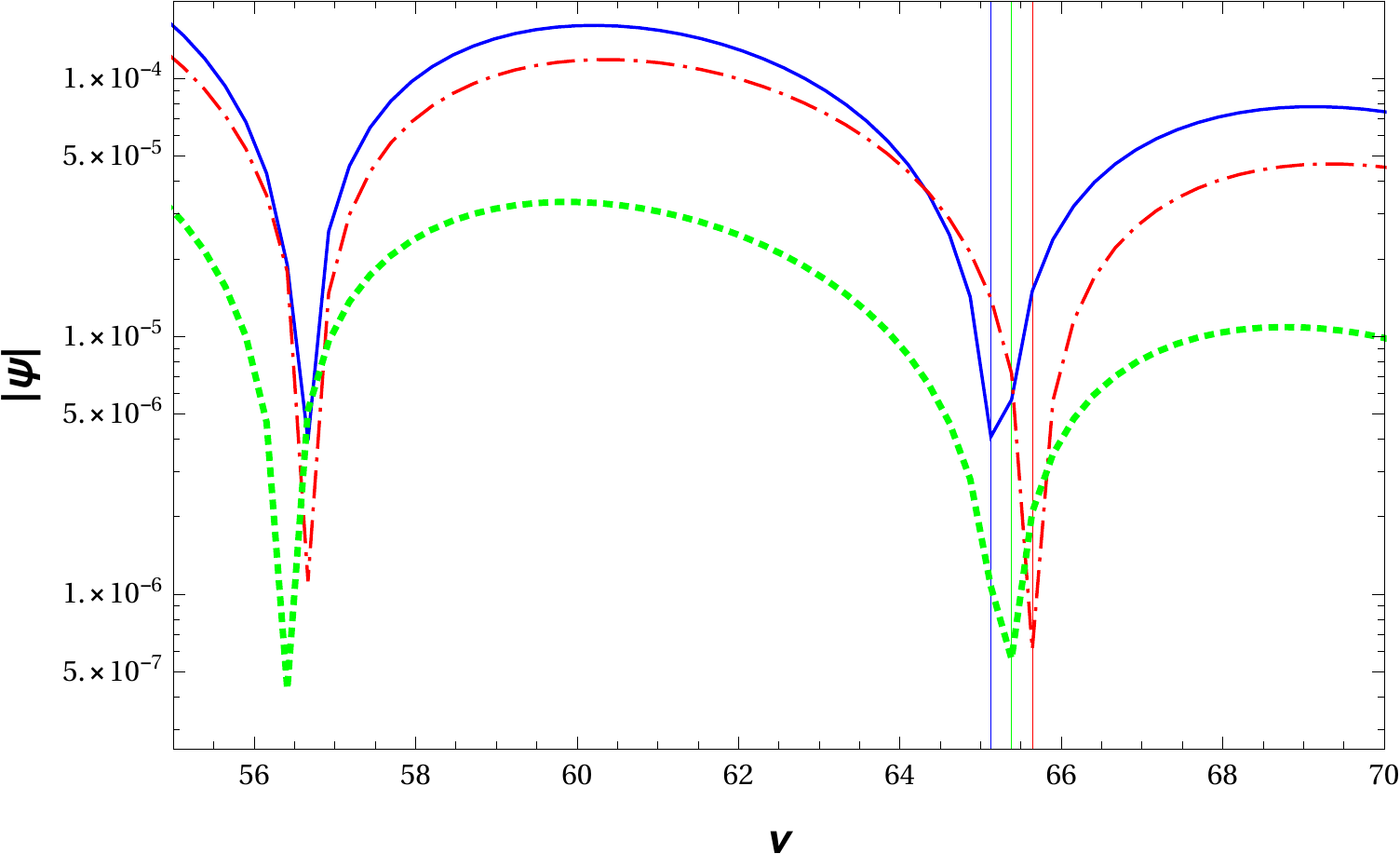} 
	\vspace*{-.3cm} \caption{\footnotesize Left panel represent the time evolution of a scalar perturbation at the small $\text{AdS}_4$ black hole horizons (semi-log graph of $|\psi|$) for $r_H=1,1.3 \ \text{and} \ 1.8 $. Right panel is a zoom of the left one. The pressure is fixed at $P=0.9P_c$\label{figtransitionsmallchi}
	}
\end{center}
\end{figure}

\begin{SCfigure}[][!ht]
	\includegraphics[width=9cm,height=6cm]{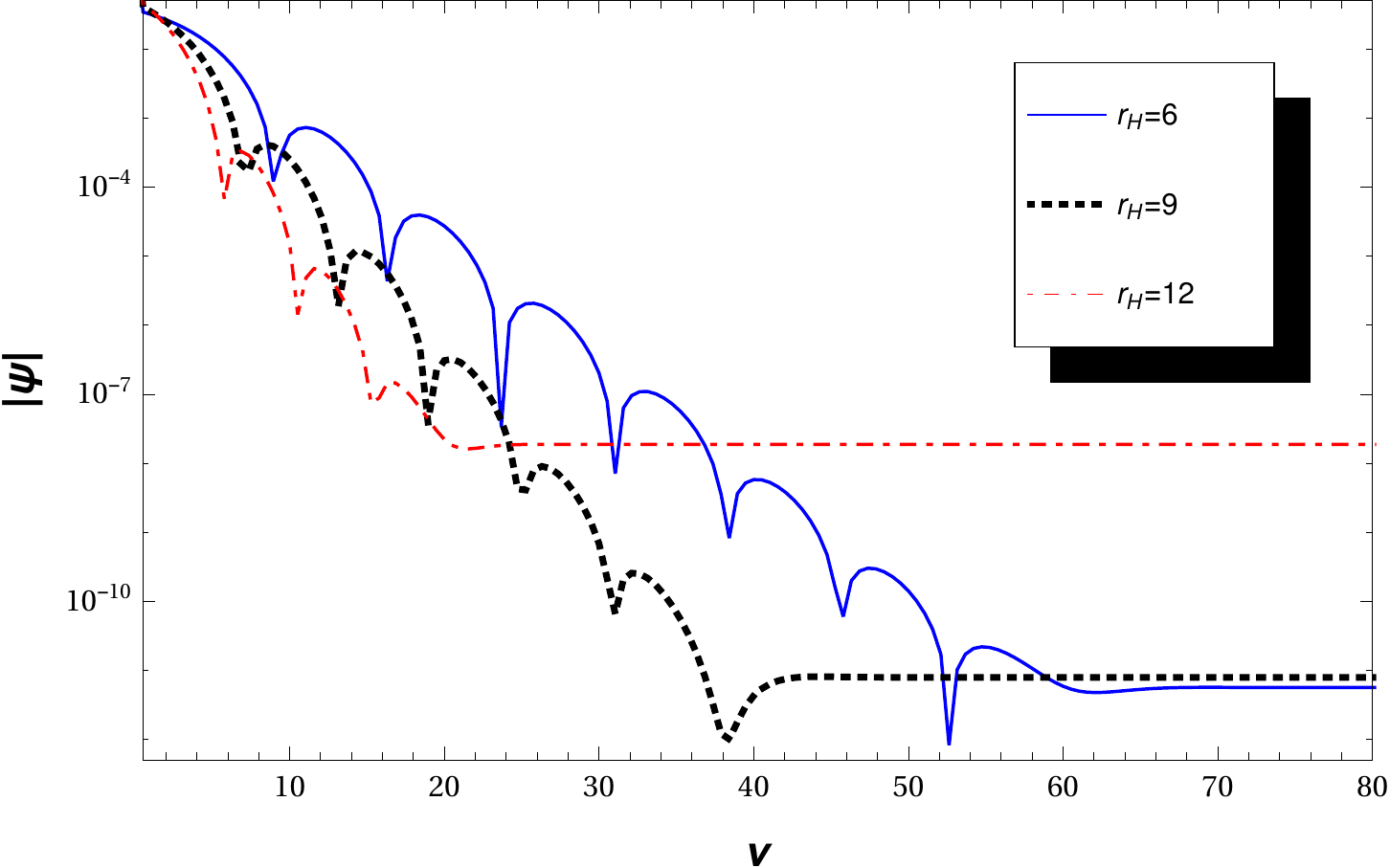}
	\vspace*{-.3cm} \caption{\footnotesize Time evolution of a scalar perturbation at the large $\text{AdS}_4$ black hole horizons (semi-log graph of $|\psi|$) for $r_H=8,9\ \text{and}\ 12 $. The pressure is fixed at $P=0.9P_c$\label{figtransitionlargechi}}
\end{SCfigure}

To compare the time domain with the results of frequency domain analysis, in figures~\ref{figtransitionsmallchi}~and~\ref{figtransitionlargechi}, we have plotted the time evolution of a scalar perturbation for small and large black hole, with $\chi=0.9$. The left panel of figure \ref{figtransitionsmallchi} is  zoom into a limited area of the right panel. Here, for the SBH, we have used  optimal values of $r_H$ ($r_{H_1}$, $r_{H_2}$ and $r_{H_3}$) chosen from figure \ref{freqrappcriti}.

In figure \ref{figtransitionsmallchi} we can see that the problem of slope change persists. When crossing over from $r_{H_1}$ to $r_{H_2}$ the perturbation decays faster and becomes less oscillating. However, from $r_{H_2}$ to $r_{H_3}$  all  the oscillations period increases  and the scalar perturbation continues to decay faster. This is again consistent with the behavior of quasinormal frequencies shown in figure \ref{freqrappcriti} since $\omega_{im}$ accounts for the damping time while  $\omega_{r}$ reflects the oscillation period. 

However, in the large black hole phase, we can see from figure \ref{figtransitionlargechi} that the perturbation presents the  same features as in  the case $P=0.4P_c$. Now one can clearly figure out that the time evolution of a perturbation in figure \ref{figtransitionlargechi} has a  similar  behavior as in green and red dash-dotted curves in figure \ref{figtransitionsmallchi}. The evolution of wave function does not show different behavior (damping time and oscillation time) in small and large black hole phases near the critical temperature, hence the QNMs are not always an efficient tool to disclose the black hole phase transition.
   

\begin{SCfigure}[][!ht]
	\includegraphics[width=9cm,height=6cm]{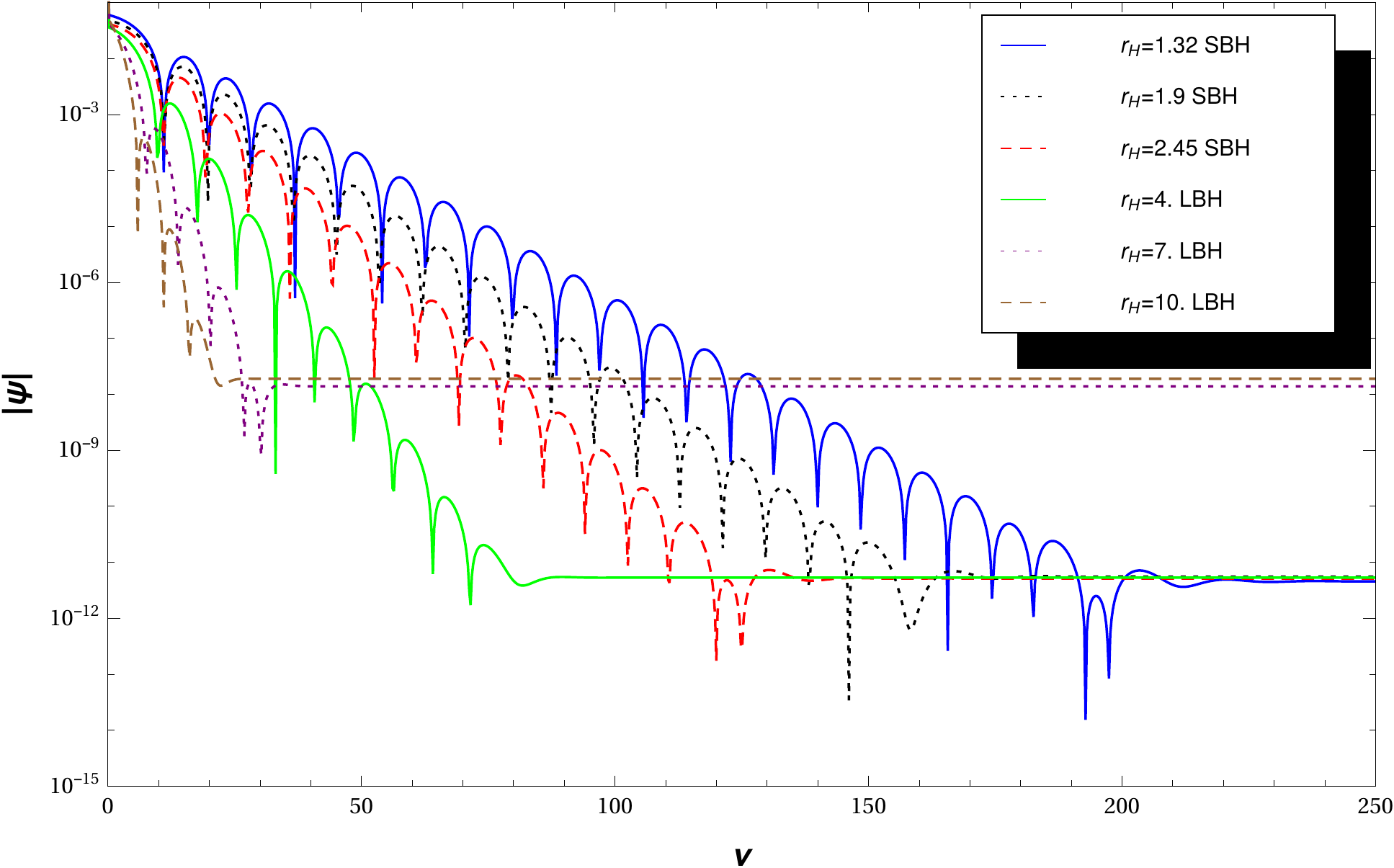}
	\vspace*{-.2cm}\caption{\footnotesize Time evolution of a scalar perturbation at the horizon of $\text{AdS}_4$ black holes  (semi-log graph of $|\psi|$) for SBH with $r_H=1.32,1.9,2.45$ and LBH with $r_H=4,7,10$. The pressure is fixed at $P=P_c$\label{tdcritical}}
  \end{SCfigure}

From \cite{our7,Liu:art_basique} we see found that, at the critical point, the quasinormal modes fail to probe the signature of the Van der Waals-like phase transitions  in the frequency domain. Now, at the end of this section, we aim to investigate  the second order phase transition signature in the time domain analysis. Here,  the situation is rather different from the first order phase transition. The quasinormal mode frequencies do not  allow us to detect the difference between the slopes of the small and large black hole phases. This  is confirmed by time domain approach in this work. Indeed,  we depict our results for 3 values of $r_H$ in each phase as illustrated in figure \ref{tdcritical}. 		
% 
% \begin{table}
% {\center \tiny
% 	\begin{tabular}{|p{0.8cm}|p{0.8cm}|p{0.8cm}p{1.15cm}|}
% 		\hline
% 		\centering{$r_{H}$} & \centering{$T$} &\centering{$\omega_{r}$}  & \hspace{9pt}$\omega_{im}$ \\%& 	\centering{$T$} & \centering{$r_{H}$} & \centering{$\omega_{r}$}  & \hspace{9pt}$\omega_{im}$\\ 
% 		\hline
% 		%\multicolumn{4}{|c}{$d=4$; $P=\frac{1}{96 \pi}$ ; $T_{c}=\frac{1}{3\sqrt{6}\pi}$}}&  	\hline
% 		 1.32 & 0.0344401 & 0.365954 & -0.117842 \\
% 		1.9 & 0.0428807 & 0.371085 & -0.147175 \\
% 		2.45 & 0.0433165 & 0.377609 & -0.184996 \\
% 		\hline\hline
% 		4. & 0.0451768 & 0.406103 & -0.299083 \\
% 		7. & 0.0575564 & 0.498377 & -0.521072 \\
% 		10. & 0.0741927 & 0.617783 & -0.742603 \\
% 		\hline 
% 		
% 	\end{tabular}
% \captionof{table}{\footnotesize  The quasinormal frequencies of the massless scalar perturbation as a function of  the black holes temperature \textcolor{red}{at the critical point.} 
% 	The upper part, above the horizontal lines is for the small black hole phase while the lower part is for the large one.}
% \label{tab2}
% }
% \end{table}
	
When the black hole (small and large) grows the damping rate increases and the perturbation decays faster while the oscillation time decreases, namely the oscillation frequencies increase. These features  revealed via the time evolution analysis confirms the behavior of quasinormal frequencies in \cite{our7,Liu:art_basique}.
%%%%%%%%%%%%%%%%%%%%%%%%%%%%%%%%%%%%%%%%%%%%%%%%%%%%%%%%%%%%%%%%%
	
\section{Conclusion}
In this paper, we studied the time evolution of a scalar perturbation around small and large $\text{RN-AdS}_4$ black holes for the purpose of probing the thermodynamic phase transition.

We found that below the critical point the scalar perturbation decays faster with increasing of the black hole size for both small and large black hole phase, but the oscillation frequencies become smaller and smaller with very slow variation.Unlike in time evolution, where the behavior analysis reveals that small and large black holes are in different phases, providing a tool to probe the BH phase transition. 

At the critical point $P=P_c$, with growing sizes of the black hole, the damping time increases and the perturbation decays faster, the oscillation frequencies increase either in small and large black hole phase. In this case the time evolution of a scalar perturbation fails to detect the $\text{AdS}_4$ black hole phase transitions confirming the result of the frequency domain analysis. 

At last, we would like to mention that it would be interesting to perform similar calculations either for  higher-dimensional RN-AdS black holes or for other black holes configurations.

\section*{Acknowledgements}
The authors are grateful to Prof. Victor Cardoso for useful discussion and comments.  This work is supported in part by the Groupement de recherche international (GDRI): Physique de l'infiniment petit et de l'infiniment grand - P2IM. %%%%%%%%%%%%%%%%%%%%%%%%%%%%%%%%%%%%%%%%%%%%%%%%%%%%%%

\end{document}